\documentclass[aps,prb,a4paper,twocolumn,floatfix,showpacs,preprintnumbers]{revtex4-1}
\usepackage{amsmath}
\usepackage{amssymb}
\usepackage{graphicx}
\usepackage{color}

\begin{document}

\title{Commensurability resonances in two-dimensional magneto-electric lateral superlattices}
\author{J. Schluck, S. Fasbender, and T. Heinzel}\email{thomas.heinzel@hhu.de}
\affiliation{Condensed Matter Physics Laboratory, Heinrich-Heine-Universit\"at, D-40225 D\"usseldorf, Germany}
\author{K. Pierz and H. W. Schumacher}
\affiliation{Physikalisch-Technische Bundesanstalt, Bundesallee 100, D-38116 Braunschweig, Germany}
\author{D. Kazazis and U. Gennser}
\affiliation{CNRS-LPN, Route de Nozay, 91960 Marcoussis, France}

\date{\today }

\begin{abstract}
Hybrid lateral superlattices composed of a square array of antidots and a periodic one-dimensional magnetic modulation are prepared in $\mathrm{Ga[Al]As}$ heterostructures. The two-dimensional electron gases exposed to these superlattices are characterized by magnetotransport experiments in vanishing average perpendicular magnetic fields. Despite the absence of closed orbits, the diagonal magnetoresistivity in the direction perpendicular to the magnetic modulation shows pronounced classical resonances. They are located at magnetic fields where snake trajectories exist which are quasi-commensurate with the antidot lattice. The diagonal magnetoresistivity in the direction of the magnetic modulation increases sharply above a threshold magnetic field and shows no fine structure. The experimental results are interpreted with the help of numerical simulations based on the semiclassical Kubo model.
\end{abstract}

\maketitle

\section{\label{sec1}INTRODUCTION}

Artificial lateral superlattices (LSLs) in two-dimensional electron gases (2DEGs) \cite{Weiss1989,Winkler1989,Weiss1991,Ensslin1990,Lorke1991} are of great interest for fundamental studies of the electron dynamics in periodic potentials. Since it is very common that the artificial lattice constants place the systems in the transition region between the quantum and the classical regime, classical, semiclassical as well as quantum descriptions are all justifiable and enable studies of the validity of these approaches including their limits. Besides the Fermi wavelength $\lambda_F$ and the electronic coherence length, the elastic mean free path is an important parameter as well, since it defines the length scale below which interaction with the LSL potential dominates over random scattering.
Many different variants of LSLs have been investigated in great depth. One-dimensional (1d) electrostatic \cite{Weiss1989,Winkler1989} and magnetic \cite{Carmona1995,Ye1995,Edmonds2001} lattices, where the modulation extends along one spatial coordinate and the structure is homogeneous along the second coordinate, show magnetoresistivity resonances that can be explained in terms of guiding center drift resonances of the cyclotron motion within a classical picture,\cite{Beenakker1989} or by miniband formation in a quantum picture. \cite{Winkler1989,Ibrahim1997,Gerhardts1996} One-dimensional magneto-electric hybrid LSLs have been studied in some experiments as well, where the strain imposed by the ferromagnetic or superconductive electrodes used to define the magnetic LSL also generates an electrostatic superlattice.\cite{Shi2002} Two-dimensional LSLs, both magnetic \cite{Ye1995,Eroms2002} and electrostatic \cite{Weiss1991,Ensslin1990,Lorke1991}, have been studied in thoroughly as well. Their classical dynamics corresponds to a mixed phase space where chaotic and regular dynamics coexist \cite{Fleischmann1992,Fleischmann1994} and causes commensurability resonances that are characteristic
for the type of Bravais superlattice employed, like square \cite{Weiss1991,Ensslin1990,Lorke1991}, rectangular \cite{Schuster1993} or hexagonal \cite{Weiss1994,Nihey1995}. Within a quantum picture, on the other hand, a fractal energy spectrum, also known as Hofstadter butterfly,\cite{Hofstadter1976,Pfannkuche1992} is seen for weak electrostatic modulation amplitudes.\cite{Geisler2004}
B-periodic oscillations on top of commensurability resonances  \cite{Weiss1993,Schuster1993,Nihey1995} can be explained within a semiclassical approach by the Aharonov-Bohm \cite{Aharonov1959} - or Altshuler-Aronov-Spivak \cite{Altshuler1981}- effect in terms of quantized motion along closed trajectories defined by the LSL potential and the magnetic field \cite{Richter1995}.\\
\begin{figure}[ht!]
\includegraphics[scale=1.0]{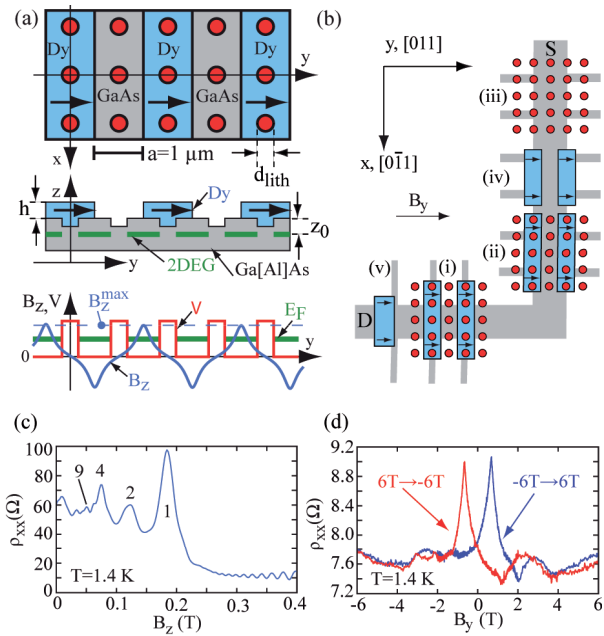}
\caption{(color online). (a) Scheme of the hybrid lateral superlattice geometry. Circular holes (red circles) that form the antidots are etched into the sample. Every second line of antidots is covered by a Dy stripe of $1\,\mathrm{\mu m}$ width. The top view of the pattern is shown in the uppermost part. In the middle, a cross section in the yz - plane at x=0 is shown. The 2DEG is depleted underneath the etched regions. The Dy stripes are magnetized in the y - direction, as indicated by the arrows. The corresponding electrostatic potential (possible strain effects are neglected) and the perpendicular magnetic field $B_z(y)$ are shown in the lowermost part. (b) Sketch of the sample layout. Two identical hybrid LSLs, (i) and (ii), are defined in an L-shaped Hall bar. The two components of the hybrid LSL, namely the antidot lattice (iii) and ferromagnetic stripes (iv) are defined separately, and the edge of a Dy pad centered in a Hall cross (v) enables Hall magnetometry. The coordinate system shows the crystallographic orientation of the Hall bar. (c) $\rho_{xx}$ of array (iii) as a function of a homogeneous perpendicular magnetic field. (d) $\rho_{xx}(B_y)$ of array (iv).}
\label{Fig1}
\end{figure}
2DEGs with very high electron mobilities \cite{Umansky1997,Roessler2010} have recently been developed into mature systems. They enable the preparation of LSLs with large lattice constants in the classical ballistic regime and facilitate the definition of novel types of LSLs with more complex unit cells. Here, the study of such a hybrid LSL, composed of a a two-dimensional, square antidot lattice and a one-dimensional magnetic array is reported. The magnetic LSL consists of approximately Lorentzian shaped peaks of alternating sign and thus has a vanishing average magnetic field. Snake trajectories, i.e., trajectories formed by the superposition of an oscillatory motion along the first direction and a motion with nonvanishing average velocity along the second direction,\cite{Nogaret2010} can become commensurate with the antidot lattice, and magnetoresistivity resonances are to be expected. Furthermore, for the magnetic modulation amplitudes applied here, closed electronic orbits are absent.\\
After the sample preparation and the experimental setup are introduced in Section \ref{sec2}, the measurements are presented in Section \ref{sec3} and interpreted with the help of numerical simulations in Section \ref{sec4}. The paper concludes with a summary and an outlook in Section \ref{sec5}.

\section{\label{sec2}SAMPLE PREPARATION AND EXPERIMENTAL SETUP}

A $\mathrm{GaAs/Al_{0.3}Ga_{0.7}As}$ heterostructure with a 2DEG $90 \,\mathrm{nm}$ below the surface is used. After a brief illumination with infrared light, the unpatterned 2DEG has a density of $3.6\times10^{15}\,\mathrm{m^{-2}}$ and a mean free path of $88\,\mathrm{\mu m}$ at liquid helium temperatures. The sample geometry is sketched in Figs. \ref{Fig1}(a) and (b). An L-shaped Hall bar, oriented parallel to the natural GaAs cleavage directions, was prepared by optical lithography. Three identical, square antidot lattices (lattice constant $a=1.0 \,\mathrm{\mu m}$) were patterned on one Hall bar by electron beam lithography and subsequent reactive ion etching. Lithographic antidot diameters of $d_{lith}=200\,\mathrm{nm}$ (sample A) as well as $d_{lith}=300\,\mathrm{nm}$ (sample B) were prepared on separate Hall bars. As a consequence of a lateral depletion length of $45\,\mathrm{nm}$ around the antidots, this corresponds to electronic diameters of $d\approx 290 \,\mathrm{nm}$ and $d\approx 390 \,\mathrm{nm}$, respectively, as measured by the Aharonov-Bohm oscillation period observed in large magnetic fields. \cite{Iye2004,supplement} Since $(a-d)/\lambda_F\approx 17$ fro sample A and $\approx 14$ for sample B, respectively, these LSLs reside well inside the classical regime. After the definition of the antidots, Dy stripes of width $a$ and a period of $2a$ were prepared on top of two antidot lattices by electron beam lithography, enabling measurements of all resistivity components in one cooldown, see Fig. \ref{Fig1}(b). The Dy stripes have a thickness of $h=250\,\mathrm{nm}$ to ensure a strong fringe field when magnetized. In sample A, they were deposited directly on the GaAs, while in sample B, a homogeneous film of $5\,\mathrm{nm}$ Cr plus $5\,\mathrm{nm}$ Au thickness was evaporated on top of the antidot lattice prior to the Dy deposition. This allows us to estimate the role of strain effects \cite{Ye1995,Beton1990} possibly induced by the Dy stripes, which are centered at the columns of antidots and aligned parallel to the x-direction.  The lateral size of the superlattices is $100\,\mathrm{\mu m}$ in longitudinal and $50\,\mathrm{\mu m}$ in  transverse direction ($100\times 25$ unit cells). For control measurements, the Hall bar furthermore contains a nominally identical magnetic stripe array without the antidots underneath, and the edge of a Dy film in a Hall cross for Hall magnetometry.\cite{Johnson1997,Cerchez2011}\\
The samples were inserted in a $\mathrm{^4He}$ gas flow cryostat with a variable temperature insert and a base temperature of $1.4\,\mathrm{K}$. The system is equipped with a magnet of $8\,\mathrm{T}$ maximum field strength. The external magnetic field $B_y$ was applied in the y - direction. It magnetizes the Dy stripes to a magnetization of $\mu_0 M (B_y)$. The 2DEG responds predominantly to the z-component of the fringe field of the Dy stripes, and we therefore neglect the influence of in-plane magnetic fields on the 2DEG throughout this paper. The magnetic field profile
$B_z(y)$ is indicated in the lowermost section of Fig. \ref{Fig1}(a). From the fringe field of a perfectly magnetized stripe, one expects \cite{Vancura2000}

\begin{widetext}
\begin{equation}
B_z(y, B_y) = \frac{\mu_0 M(B_y)}{4\pi}\sum\limits_{j=0}^{N-1}\ln \left(\frac{A^{-}}{A^{+}}\right); A^{\pm}= \frac{[y-a(2j\mp \frac{1}{2})]^2+z_0^2}{[y-a(2j\mp \frac{1}{2})]^2+(z_0+h)^2}
\label{eq1}
\end{equation}
\end{widetext}

where $z_0$ is the distance between the 2DEG and the bottom of the Dy film, j is an integer and N denotes the total number of Dy stripes. This magnetic profile has peaks of alternating sign with amplitude  $B_z^{max}(B_y)\equiv \left| B_z(y=\left[2j-\frac{1}{2}\right]a, B_y)\right|$. The maximum  magnetization of our Dy films is $\mu_0 M\approx 2.7 \,\mathrm{T}$ for $B_y>5\,\mathrm{T}$, corresponding to an upper limit of $B_z^{max}\approx 480\,\mathrm{mT}$. The coercive magnetic field is $B_c=670\,\mathrm{mT}$. The resistivity components $\rho_{ij}(B)$ with $i,j \epsilon \{x,y\}$ were determined by applying an AC current of $100\,\mathrm{nA}$ with a frequency of $17.7\,\mathrm{Hz}$ from source S to drain D, see Fig. \ref{Fig1}(b), and by measuring the electrostatic potentials in x- and y-direction at voltage probes with a lock-in amplifier.\\

\section{\label{sec3}EXPERIMENTAL RESULTS}

\begin{figure}[ht!]
\includegraphics[scale=1.0]{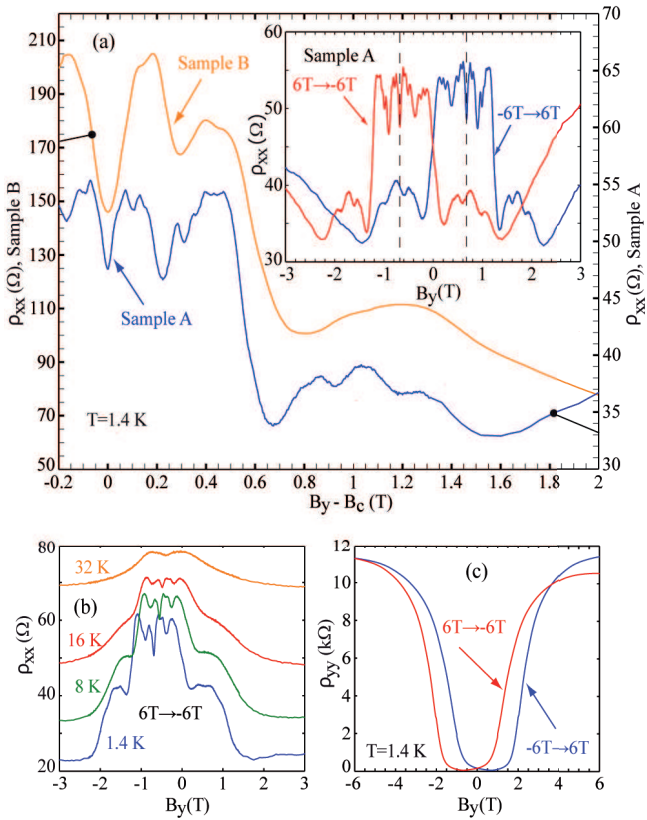}
\caption{(color online). (a) Magnetoresistivities $\rho_{xx} (B_y-B_c)$ of the hybrid superlattices of samples A and B, shown for the up-sweeps (increasing $B_y$). The inset shows the hysteretic behavior, exemplified for sample A. The dashed lines indicate $B_c$. (b) Temperature dependence of $\rho_{xx} (B_y)$ for sample A, as observed in a different cooldown. Only the down-sweep direction is shown for clarity. The measurement of $\rho_{yy}(B_y)$ for sample A is reproduced in (c).}
\label{Fig2}
\end{figure}

In Figure \ref{Fig1} (c) and (d), the magnetoresistivities of the LSL components of sample A for the two hybrid LSL components, namely the square antidot lattice (iii) and the array of magnetic stripes (iv), respectively, are reproduced. The antidot lattice reveals the well-known commensurability resonances with resistivity maxima at perpendicular magnetic fields where the cyclotron orbit is commensurate with one, two, four or nine enclosed antidots.\cite{Weiss1991,Ensslin1990,Lorke1991} For $B_z>250\,\mathrm{mT}$, Shubnikov - de Haas oscillations set in. $\rho_{xx}(B_y)$ of the Dy stripes shows a peak centered at $B_c$ and some weak features at larger magnetizations. This type of magnetoresistivity of magnetic stripe arrays in in-plane magnetic fields has been studied theoretically,\cite{Ibrahim1995} while to the best of our knowledge, experiments have been reported only in related configurations.\cite{Nogaret2011} Numerical simulations based on the classical Kubo formalism (see below for details) give a weak, positive magnetoresistivity without fine structure,\cite{supplement} as measured for $|B_y|\gtrsim 4\,\mathrm{T}$. This indicates that the peak at $B_c$ is not an intrinsic classical property of the magnetic profile itself, and we tentatively attribute it to the frequently observed and still not fully understood negative colossal magnetoresistance in high-mobility 2DEGs,\cite{Bockhorn2011,Hatke2011,Shi2014} which is beyond our focus here, possibly in combination with other effects like weak localization. The strength of this feature depends on the cooldown cycle. It should be emphasized that $\rho_{xx}$ of the 2DEG underneath the Dy array is constant over the full scan range within $\pm 0.8\,\mathrm{\Omega}$. For the following, this contribution can therefore be neglected.

The diagonal magnetoresistivities $\rho_{xx}(B_y)$ and $\rho_{yy}(B_y)$ of the hybrid LSLs  are reproduced in Fig. \ref{Fig2}, and we first focus on $\rho_{xx}(B_y)$ as observed on samples A and B (a). As $B_y$ is detuned away from $B_c$, a positive magnetoresistivity is observed. As $B_y$ is further increased, two peaks are seen, separated by a pronounced minimum. Around $B_y-B_c\approx 600\,\mathrm{mT}$, a decrease of $\rho_{xx}$ by roughly a factor of 2 is seen, followed by a broad maximum that extends up to $B_y-B_c\approx 2\,\mathrm{T}$. These most prominent features appear in sample B at somewhat larger magnetic fields than in sample A. Also, even though the positive magnetoresistance is less pronounced in sample A than in sample B, sample A shows clear additional finer structures, some of which are also adumbrated in $\rho_{xx}(B_y)$ of sample B. These differences can be traced back to the Cr/Au electrode present in sample B, as will be discussed below in more detail. In the following, we focus on sample A. In the inset of Fig. \ref{Fig2} (a), the hysteretic behavior of $\rho_{xx}(B_y)$ is reproduced. The features are fairly symmetric about $B_c$, while the symmetry of the up-sweep to the down-sweep about $B_y=0$ is close to perfect. This behavior indicates that the magnetization of the Dy stripes is not perfectly antisymmetric about $B_y=B_c$ (see below).  These magnetoresistivity features show a weak temperature dependence, see Fig. \ref{Fig2} (b), and the most pronounced ones remain visible up to $T\approx 16\,\mathrm{K}$. This suggests that they should be interpretable within a classical picture. They are furthermore superimposed to a slowly varying negative magnetoresistivity that extends to $|B_y-B_c|\approx 1.6\,\mathrm{T}$, after which it increases slightly. This background depends somewhat on the cooldown cycle. The strong positive magnetoresistivity in a narrow interval around $B_c$ is still clearly visible at $32\,\mathrm{K}$, and behaves similarly to that one observed in two-dimensional antidot lattices, see also Fig. \ref{Fig1} (c). It is due to a $B_z$ - induced increase in scattering at the antidots and is of no further interest here.\\

A smooth increase of $\rho_{yy} (B_y)$ is observed as $B_y$ is driven away from $B_c$. A sharp increase sets in for $|B_y-B_c|\approx 1.6\,\mathrm{T}$ and stops for $|B_y-B_c|\approx 3.5 \,\mathrm{T}$, see Fig. \ref{Fig2} (c). The shape of $\rho_{yy} (B_y)$ strongly resembles that one observed for single magnetic barriers,\cite{Vancura2000} as well as magnetic barriers in series of alternating polarity. \cite{Kubrak1999} Within a classical picture, the increasing amplitude of $B_z(y)$ reflects an increasing fraction of the incident electrons that gets reflected at the magnetic barrier. Above a threshold amplitude of $B_z(y)$, electrons can only pass the barrier via $\vec{E}\times\vec{B}$ drift at the edges of the Hall bar, or by scattering events inside the magnetic barrier.\cite{Cerchez2007} These effects cause a saturation of $\rho_{yy}$ at large $B_z^{max}$. Since our ferromagnetic array represents an array of magnetic double barriers in series, \cite{Kubrak1999}, $\rho_{yy} (B_y)$ can thus be qualitatively understood in terms of the properties of magnetic double barriers with the antidots acting as scatterers, \cite{supplement} and is not a unique signature of the the hybrid lattice. The onset of the sharp increase of $\rho_{yy}$ furthermore correlates with the end of the negative magnetoresistivity in x-direction. Comparison of $\rho_{xx}(B_y)$ to $\rho_{yy}(B_y)$ reveals that the transport at large magnetic fields is highly anisotropic. For example, for $|B_y-B_c|= 2.5\,\mathrm{T}$, the ratio $\rho_{yy}/\rho_{xx}$ reaches a value of $\approx 230$. This suggests that for sufficiently large magnetization of the Dy stripes, the electrons are guided along the x-direction by the magnetic modulation, while crossing the magnetic walls is highly unlikely.

The off-diagonal components of the magnetoresistivity tensor were measured as well.\cite{supplement} Since the average perpendicular magnetic field is zero, they vanish to a good approximation in the magnetic field range where the resonances in $\rho_{xx}(B_y)$ appear and are thus not very helpful for their interpretation.

\section{\label{sec4}MODEL CALCULATION AND DISCUSSION}

A coarse estimation, to be substantiated below, reveals that for $B_z^{max}\lesssim 500\,\mathrm{mT}$, $B_z(y)$ is too weak to generate closed cyclotron-type orbits. Therefore, the magnetoresistivity resonances must originate from open trajectories. This situation is quite different in comparison to antidot lattices in homogeneous magnetic fields where closed orbits, runaway trajectories and chaotic orbits coexist and all contribute to the magnetoresistivity with a magnetic field-dependent weight.\cite{Zozoulenko1997} Open cycloid orbits are absent as well in the interval where the resonances appear, and it is therefore expected that snake trajectories play an important role, the most obvious type of which is centered at the roots of $B_z(y) $ and runs along columns of antidots in x-direction. Since this is a classical picture and moreover the most pronounced features of the magnetoresistivity show a weak temperature dependence, it appears plausible to model them using the classical Kubo formalism. The code we use has been presented in detail elsewhere \cite{Meckler2005} and is therefore only briefly sketched here. We show the simulations for the parameters of sample A. Electrons are injected at random locations inside a unit cell of the LSL. They initially move in random directions with their Fermi velocity of $v_F=2.6\times 10^{5}\,\mathrm{m/s}$. The incremental change of the direction of motion by the inhomogeneous magnetic field given by Eq. \eqref{eq1} is calculated with a step width of $2\,\mathrm{nm}$, and specular reflection at the antidots with $d=290\,\mathrm{nm}$ is assumed. Furthermore, $z_0=90\,\mathrm{nm}$ is used, and we assume that the antidot potential is hard-wall, as justified by the large $a/d$ ratio. The simulations are carried out for zero temperature. From the simulated diffusion tensor obtained via the Kubo formula, the  magnetoresistivity components are obtained via the Einstein relation for a degenerate 2DEG.

\begin{figure}[ht!]
\includegraphics[scale=1.0]{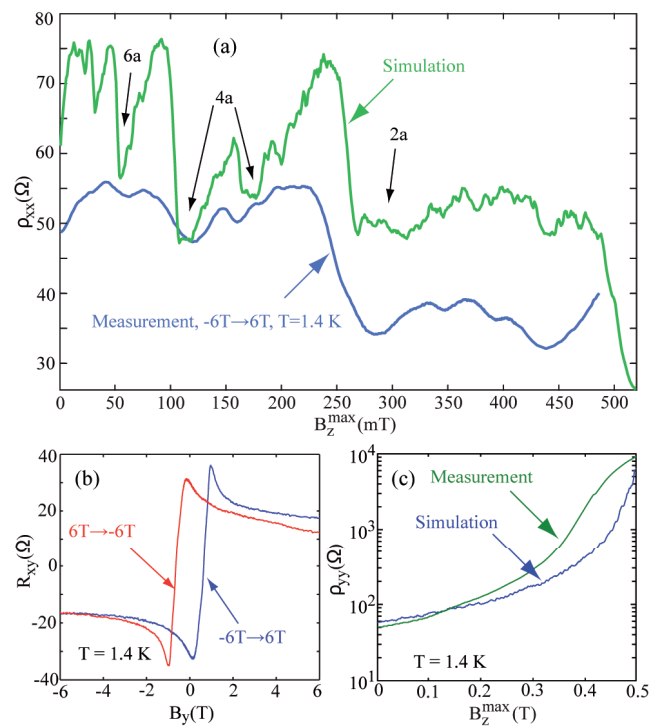}
\caption{(color online). (a) $\rho_{xx}(B_z^{max})$ as simulated within the Kubo model, plotted as a function of the maximum of $B_z$. The period of the commensurate snake trajectories is indicated at the most prominent resistivity minima. The experimental trace of sample A has been scaled to $B_z$ with the help of the magnetization trace, as obtained from the Hall resistance of one Dy edge centered inside a Hall cross (b). (c) Simulation results for $\rho_{yy}(B_z^{max})$ in comparison to the scaled experimental data of sample A.}
\label{Fig3}
\end{figure}

Figure \ref{Fig3} (a) shows the simulated magnetoresistivity $\rho_{xx}$  as a function of the maximum of the perpendicular magnetic field $B_z^{max}$, see also Figs. \ref{Fig1} (a) and (b).

As in the experiment, several features in $\rho_{xx}$ are observed. Close to $B_z^{max}< 250\,\mathrm{mT}$, a positive magnetoresistivity is present. For $B_z^{max}< 250\,\mathrm{mT}$, a series of resistivity minima at $B_z^{max}=32\,\mathrm{mT}, 53\,\mathrm{mT}$ and $\approx 110\,\mathrm{mT}$ is visible. A clear but weaker additional minimum is visible at $B_z^{max}\approx 170\,\mathrm{mT}$. Above a sharp decrease of $\rho_{xx}$ at $B_z^{max}\approx 260\,\mathrm{mT}$, a broad minimum around $280\,\mathrm{mT}$ is present, followed by some weakly pronounced maxima and minima. Finally, another sharp decrease of $\rho_{xx}$ around $B_z^{max}=500\,\mathrm{mT}$ is observed.  A direct comparison with the measurements requires knowledge of the transformation function  $\mu_0 M (B_y)$. Conceptually, it can be determined by Hall magnetometry of the stripe array on top of Hall crosses well inside the diffusive regime. In the ballistic or quasi-ballistic regime, the Hall voltage translates into the magnetization by nontrivial correction factors, \cite{Geim1997b,Cerchez2011} the detailed discussion of which is beyond our scope here. Since such an estimation would still assume perfect, mono-domain magnetization of the Dy stripes as well as a certain shape of the fringe field, some uncertainty would remain.  Therefore, in order to estimate $\mu_0 M(B_y)$, we restricted ourselves to Hall magnetometry of the edge of a Dy film, prepared in the same process step as the magnetic lattice. The measured Hall voltage as a function of $B_y$, reproduced in Fig. \ref{Fig3} (b), shows a marked peak where the average cyclotron diameter equals the width of the voltage probe. The decrease of the Hall voltage at larger magnetic fields originates from ballistic effects.\cite{Cerchez2011} The asymmetry of the Hall voltage furthermore indicates that the magnetization of the film is not perfect.  Therefore, we compare the measured data to the simulations by scaling it with an approximated function $\mu_0M(B_y)$, obtained numerically along the lines of Ref. \onlinecite{Cerchez2011}, where $\mu_0M$ is roughly proportional to $B_y$ for $|B_y-B_c|<600\,\mathrm{mT}$ and depends only weakly on $B_y$ for larger applied magnetic fields. This analysis of the Hall magnetometry indicates a saturation magnetization for the Dy stripes of $\approx 2.7\,\mathrm{T}$, and $B_c=670\,\mathrm{mT}$ can be read out directly. The data measured at sample A in the up-sweep for $B_y>B_c$ in Fig. \ref{Fig2} (a) are scaled accordingly and replotted in Fig. \ref{Fig3} (a) as a function of $B_z^{max}$, which allows a more direct comparison to the simulations. \\
Even though the simulated function $\rho_{xx}(B_z^{max})$ deviates from the experimental trace in several aspects, the most prominent features are reproduced qualitatively, namely the positive magnetoresistivity around $B_z^{max}=0$, minima close to $B_z^{max}=53\,\mathrm{mT}, 110\,\mathrm{mT}, 170\,\mathrm{mT}$ and $280\,\mathrm{mT}$, the decrease of $\rho_{xx}$ at $B_z^{max}\approx 260\,\mathrm{mT}$, and some weakly pronounced maxima and minima at larger magnetic fields. The sharp decrease of of $\rho_{xx}$ around $B_z^{max}=500\,\mathrm{mT}$ is not observed experimentally, most likely because our fringe fields are too weak.

The simulation of $\rho_{yy}(B_z^{max})$ is compared to the scaled experimental data in Fig. \ref{Fig3} (c). Very good agreement is found for $B_z^{max}\leq 0.3\,\mathrm{T}$, while the strong increase of the resistivity around $B_z^{max}\approx 0.4\,\mathrm{T}$ is reproduced as well, though shifted to slightly higher magnetic fields. Further simulations \cite{supplement} show that the presence  of the antidots does influence $\rho_{yy}(B_z^{max})$ to some extent, but the overall behavior is dominated by the magnetic barriers and is not an effect of the hybrid superlattice.\\

We proceed by interpreting the magnetoresistivity features in terms of the electron dynamics which determines the components of the magnetoconductivity tensor.\cite{supplement} The off-diagonal elements $\sigma_{xy}$ and $\sigma_{yx}$ are approximately independent of $B_z^{max}$ and of the order of $0.1\,\mathrm{mS}$. $\sigma_{yy}$ decreases from $18\,\mathrm{mS}$ at $B_z^{max}=0$ to almost zero at $B_z^{max}\approx 0.5\,\mathrm{T}$. Only $\sigma_{xx}$ shows resonances as $B_z^{max}$ is changed. This implies that $\rho_{xx}\approx 1/\sigma_{xx}$ and $\rho_{yy}\approx 1/\sigma_{yy}$, while $\rho_{xy}(B_y)\approx \sigma_{xy}/(\sigma_{xx}\sigma_{yy})$. The sharp increase of $\rho_{xy}$ (see Ref. \onlinecite{supplement}) and $\rho_{yy}$  at $B_z^{max}=0.5\,\mathrm{T}$ has thus its origin in the strongly suppressed diagonal conductivity in y-direction.

\begin{figure}[ht!]
\includegraphics[scale=1.0]{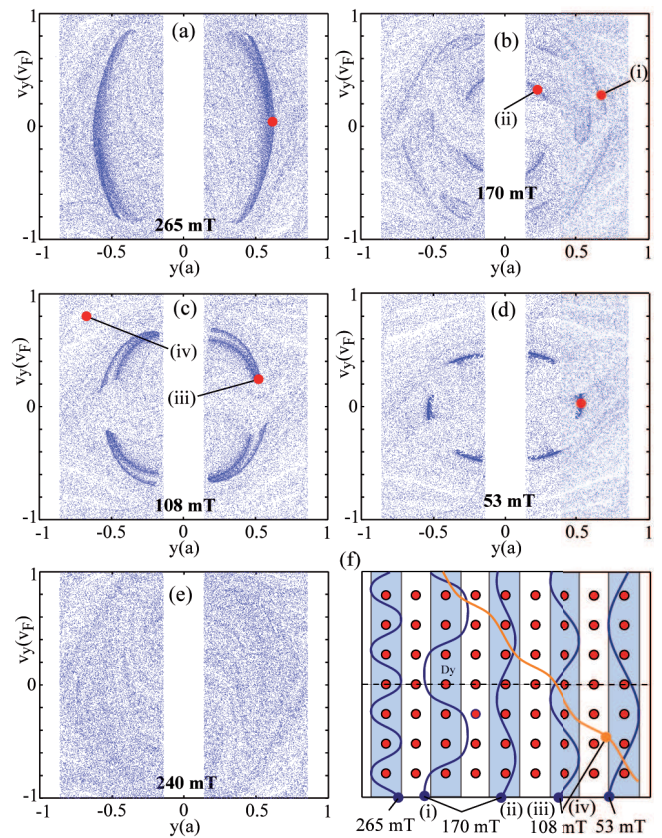}
\caption{(color online). Poincar$\mathrm{\acute{e}}$ sections for various values of $B_z^{max}$ (a-e). Some characteristic trajectories are shown in (f), the initial conditions of which are indicated by full circles in the corresponding Poincar$\mathrm{\acute{e}}$ sections.}
\label{Fig4}
\end{figure}

A deeper insight into the underlying electron dynamics can be gained by looking at characteristic electron trajectories. They can be identified with the help of Poincar$\mathrm{\acute{e}}$ sections, which illustrate the dynamics of the electrons by their coordinates in a $(i,v_j)$ cross section of the phase space ($i = x,y$ and $v_j, j=x,y$ denote the position and velocity coordinates, respectively). We start with a discussion of the minima of $\rho_{xx}$ at smaller magnetic fields, $B_z^{max}< 280\,\mathrm{mT}$. Each dot in Figs. \ref{Fig4} (a-e) represents the coordinates of an electron passing with $v_x>0$ through one of the $(y, v_y)$ - planes located at at $x= m a$, where $m$ is an integer. The Poincar$\mathrm{\acute{e}}$ section for $B_z^{max}=265\,\mathrm{mT}$  (a) shows a pronounced accumulation of the electrons in a semicircle-like structure that extends over 85\% of possible $v_y$ components. This region hosts quasi-commensurate snake trajectories with a wavelength very close to 2a. They run parallel to the magnetic stripes, as illustrated by the sample trajectories shown Fig. \ref{Fig4} (f), and typically get scattered at the antidots after less than 30 snake periods. Likewise, the Poincar$\mathrm{\acute{e}}$ sections for the minima of $\rho_{xx}$ at $B_z^{max}=170\,\mathrm{mT}$ (b), $108\,\mathrm{mT}$ (c), and at $53\,\mathrm{mT}$ (d) reveal that here, quasi-commensurate snake trajectories of various periodicity exist. They extend along the x - direction, and their weight decreases as the magnetic field is decreased, which correlates with the magnitude of the corresponding resistivity dips. Outside the resistivity minima, such an accumulation of electrons in snake trajectories is not seen in the Poincar$\mathrm{\acute{e}}$ sections, as illustrated for $B_z^{max}=240\,\mathrm{mT}$ in Fig. \ref{Fig4} (e).\\
In addition, snake orbits exist which run at an angle $\neq 0$ to the x-direction, as exemplified in Fig. \ref{Fig4} (f). In the Poincar$\mathrm{\acute{e}}$ sections, such trajectories form white regions, since the electrons do not return to the column in which the electrons start. They can be found over the whole interval where resonances are observed, and we do not find a correlation between their weight in the Poincar$\mathrm{\acute{e}}$ section and the magnetoresistivity. We furthermore emphasize that, as anticipated above, closed orbits are absent. It thus emerges that the minima of $\rho_{xx}(B_y)$ for $B_z^{max}\lesssim 280\,\mathrm{mT}$ correlate with the presence of quasi-commensurate snake trajectories that run parallel to the magnetic stripes for many antidot periods, while snake trajectories running in other directions do not show such a correlation. Both the depletion and accumulation regions of the Poincar$\mathrm{\acute{e}}$ sections are embedded in an approximately homogeneously filled background, which is due to  electrons that move in snake orbits as well, but experience frequent scattering at the antidots. Typically, such trajectories complete no more that two snake periods before they get scattered. \cite{supplement} We note that both the accumulation and the depletion regions contain mostly not perfectly periodic trajectories and are thus chaotic as well. Regular orbits should exist inside the accumulation regions, but we have been unable to identify such points in the Poincar$\mathrm{\acute{e}}$ sections, which indicates that the regular regions have a very small volume. The composition of the phase space of this hybrid LSL is thus different to that one of antidot lattices where disjunct, extended regular and chaotic regions coexist.

It is remarkable that adjacent resistivity minima sometimes correlate with accumulations of snake orbits of the same periodicity. For example, the minima at $B_z^{max}=170\,\mathrm{mT}$ and at $108\,\mathrm{mT}$ both correlate with the accumulation of snake trajectories with a period close to 4a. While the snake trajectories that belong to the pronounced minimum of $\rho_{xx}$ at $108\,\mathrm{mT}$ remain commensurate over a relatively large interval of magnetic fields and initial conditions, those found at the weak minimum at $170\,\mathrm{mT}$, like the two shown in (f) with their location indicated in the Poincar$\mathrm{\acute{e}}$ section in (b), are more fragile.

\begin{figure}[ht!]
\includegraphics[scale=1.0]{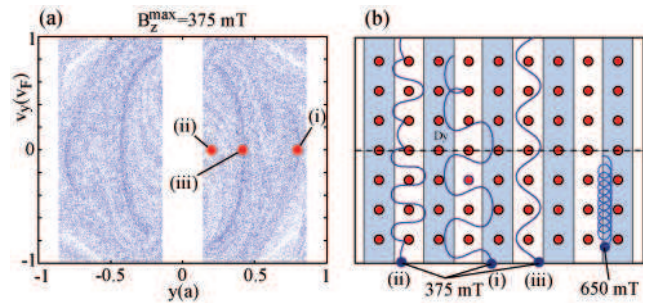}
\caption{(color online). (a) Poincar$\mathrm{\acute{e}}$ section  for $B_z^{max}= 375\,\mathrm{mT}$. (b) Some typical trajectories at $B_z^{max}= 375\,\mathrm{mT}$ with initial conditions as indicated by full circles in (a), and a cycloid trajectory for $B_z^{max}= 650\,\mathrm{mT}$.}
\label{Fig5}
\end{figure}

For $B_z^{max}> 280\,\mathrm{mT}$, the simulation shows a series of weakly pronounced features that end with a strong decrease of $\rho_{xx}$ at $B_z^{max}\approx 500\,\mathrm{mT}$. Qualitatively similar features are observed experimentally for sample A and can be only guessed for sample B. In this interval, the Poincar$\mathrm{\acute{e}}$ sections show a rich pattern of accumulation regions, together with a few depleted areas, see Fig. \ref{Fig5} (a). This pattern evolves smoothly as a function of $B_z^{max}$ without changing its qualitative appearance. The snake trajectories in this interval have a period close to 2a and show only a few oscillations before they get scattered. We also find occasional trajectory segments of skipping orbits, see Fig. \ref{Fig5} (b). Thus, the magnetoresistivity features in the interval $280\,\mathrm{mT}<B_z^{max}< 500\,\mathrm{mT}$ do not correlate in a straightforward way with characteristic trajectories.

The limit of large magnetic fields is characterized by $B_z^{max}\approx 500\,\mathrm{mT}$. It is not experimentally accessible in our samples. The simulations suggest that the decrease of $\rho_{xx}$ originates from the formation of cycloid trajectories which drift along the magnetic field peaks. An example of such a trajectory is shown in Fig. \ref{Fig5} (b). For $B_z^{max}> 520\,\mathrm{mT}$, cycloid orbits exist that never hit an antidot. Therefore, highly conductive channels in x-direction are formed.

We conclude this section by discussing possible reasons for the differences observed between sample A and sample B, as well as for the deviations between the simulations and the experiments. The most prominent features in sample B appear at higher magnetic fields than their respective counterparts in sample A. With the support of the corresponding numerical simulations, this can be traced back to the larger distance of the Dy stripes to the 2DEG due to the Cr/Au  layer in between, which makes a higher magnetization necessary to achieve a fringe field of the same magnitude at the depth of the 2DEG. Also, the less prominent features observed at sample A are suppressed in sample B. This may be due to the larger antidot diameters in sample B which is known to smear out commensurability resonances.\cite{Weiss1991}. Also, gating of high mobility heterostructures can decrease the mobility.\cite{Roessler2010}

The simulated amplitudes of the commensurability oscillations are furthermore stronger than the measured ones. We attribute this partly to the deviations of the real magnetic field profile $B_z(y)$ from the simulated one, which is to be expected from the asymmetric magnetization characteristics of the Dy film. Deviations from the assumed hard-wall potential may deform the trajectories, thereby weakening the resonances. Another possible reason are piezoelectric effects due to strain imposed by the Dy stripes, which could modulate the electron density and the mobility for our crystalligraphic orientation of the Hall bars. This effect has been reported in the literature to get attenuated by depositing the stripes on top of a homogeneous metallic layer.\cite{Ye1995} Therefore, by comparing the measurements of sample A with those of sample B, we conclude that if strain effects were relevant, they would generate additional fine structure rather than smearing them out. To further elucidate this issue, we have performed numerical simulations as described above, with an additional electrostatic potential of a cosine shape in y direction with the period of the magnetic stripes and a rather strong amplitude of $1\,\mathrm{meV}$, in accordance to typical values found in the literature.\cite{Beton1990,Larkin1997} Somewhat surprisingly, we do not find significant deviations of the resistivity from the unmodulated case (not shown) and therefore conclude that the magnetic field gradient dominates over electrostatic effects in the regime where the resonances are observed. Strain effects thus do not play a prominent role. Also, the simulation neglects finite size effects. For example, a magnetic barrier close to the Hall bar edges induces $\vec{E}\times\vec{B}$ drift, and electron scattering at the Hall bar edges may provide additional conductance channels. Finally, at the large in-plane magnetic fields present in our implementation, magnetic mass effects can deform the snake trajectories to a small extent.\cite{Sotomayor2002}\\

\section{\label{sec5}SUMMARY AND CONCLUSIONS}

Hybrid magneto-electric lateral superlattices composed of a two-dimensional antidot array and a one-dimensional magnetic modulation have been defined in high-mobility two-dimensional electron gases and studied by transport experiments in a configuration with vanishing average perpendicular magnetic field.  Despite the absence of closed trajectories, pronounced classical magnetoresistivity resonances have been observed. The magnetoresistivity minima correlate with the accumulation of electrons in snake trajectories, as observed in  Poincar$\mathrm{\acute{e}}$ sections, that are quasi-commensurate with the antidot lattice and oriented along the direction in which the magnetic field is homogeneous. Snake trajectories running in other directions are present as well, but their appearance does not correlate with the resistivity minima. The Poincar$\mathrm{\acute{e}}$ sections do not show extended regular islands.  We hope that these rsults will trigger quantum simulations of this system which should be able to interpret the magnetoresistivity resonances on a more fundamental level. The longitudinal magnetoresistivity is furthermore strongly anisotropic, with resistivity ratios above 200 for large magnetic fields. To a good approximation, however, the magnetoresistance in the direction perpendicular to the magnetic stripes can be understood as a resistance of magnetic barriers in series and does not reveal superlattice-specific properties. Further experiments may comprise the application of additional homogeneous perpendicular magnetic fields, a more detailed study of $\rho_{yy}$, the interaction of the electrons in snake trajectories with resonant electromagnetic radiation, or magnetic mass effects.

The authors would like to thank HHU D\"usseldorf for financial support.

%\bibliography{E:/Research/Manuscripts/2015_manuscripts/LSHeinzel_2015_01_27}
%merlin.mbs apsrev4-1.bst 2010-07-25 4.21a (PWD, AO, DPC) hacked
%Control: key (0)
%Control: author (8) initials jnrlst
%Control: editor formatted (1) identically to author
%Control: production of article title (-1) disabled
%Control: page (0) single
%Control: year (1) truncated
%Control: production of eprint (0) enabled
%

\end{document}